\documentclass[twocolumn,amsmath,amssymb,superscriptaddress,showpacs,preprintnumbers,prl,footinbib,aps,10pt]{revtex4-1}

\usepackage[T1]{fontenc}
\usepackage{textcomp}
\usepackage{txfonts}
\usepackage{bm}
\usepackage{graphicx,color,epsfig}
\usepackage{xspace}
\usepackage{stmaryrd}
\usepackage{dcolumn}   
\def\bbbc{{\mathchoice {\setbox0=\hbox{$\displaystyle\rm C$}\hbox{\hbox
to0pt{\kern0.4\wd0\vrule height0.9\ht0\hss}\box0}}
{\setbox0=\hbox{$\textstyle\rm C$}\hbox{\hbox
to0pt{\kern0.4\wd0\vrule height0.9\ht0\hss}\box0}}
{\setbox0=\hbox{$\scriptstyle\rm C$}\hbox{\hbox
to0pt{\kern0.4\wd0\vrule height0.9\ht0\hss}\box0}}
{\setbox0=\hbox{$\scriptscriptstyle\rm C$}\hbox{\hbox
to0pt{\kern0.4\wd0\vrule height0.9\ht0\hss}\box0}}}}

\DeclareFontFamily{OT1}{pzc}{}
\DeclareFontShape{OT1}{pzc}{m}{it}{<-> s * [1.2] pzcmi7t}{}
\DeclareMathAlphabet{\mathpzc}{OT1}{pzc}{m}{it}
\newcommand{\ham}{\mathpzc{H}}

\usepackage{pict2e}

\newsavebox{\dimerIIIA}
\savebox{\dimerIIIA}[10.5pt][s]{%
  \setlength{\unitlength}{2.pt}
  \begin{picture}(3.464,4)(0.3,0.6)
    \linethickness{1.6pt}
    \put(1.732,4){\line(1.732,-1){1.732}}
    \put(1.732,0){\line(1.732,1){1.732}}
    \put(0,1){\line(0,1){2}}
    \linethickness{0.05pt}
    \put(1.732,4){\line(-1.732,-1){1.732}}
    \put(1.732,0){\line(-1.732,1){1.732}}
    \put(3.464,1){\line(0,1){2}}
  \end{picture}
  \setlength{\unitlength}{1pt}
}

\newsavebox{\dimerIIIB}
\savebox{\dimerIIIB}[10.5pt][s]{%
  \setlength{\unitlength}{2.pt}
  \begin{picture}(3.464,4)(0.3,0.6)
    \linethickness{1.6pt}
    \put(1.732,4){\line(-1.732,-1){1.732}}
    \put(1.732,0){\line(-1.732,1){1.732}}
    \put(3.464,1){\line(0,1){2}}
    \linethickness{0.05pt}
    \put(1.732,4){\line(1.732,-1){1.732}}
    \put(1.732,0){\line(1.732,1){1.732}}
    \put(0,1){\line(0,1){2}}
  \end{picture}
  \setlength{\unitlength}{1pt}
}

\begin{document}

\title{Stiffness from Disorder in Triangular-Lattie Ising Thin Films }

\author{Shi-Zeng Lin}
\affiliation{
  Theoretical Division, T-4 and CNLS, Los Alamos National Laboratory, 
  Los Alamos, New Mexico 87545, USA
}
\author{Yoshitomo Kamiya}
\affiliation{
  iTHES Research Group and Condensed Matter Theory Laboratory, RIKEN,
  Wako, Saitama 351-0198, Japan
}
\affiliation{
  Theoretical Division, T-4 and CNLS, Los Alamos National Laboratory, 
  Los Alamos, New Mexico 87545, USA
}
\author{Gia-Wei Chern}
\affiliation{
  Theoretical Division, T-4 and CNLS, Los Alamos National Laboratory, 
  Los Alamos, New Mexico 87545, USA
} 
\author{Cristian D. Batista}
\affiliation{
  Theoretical Division, T-4 and CNLS, Los Alamos National Laboratory, 
  Los Alamos, New Mexico 87545, USA
} 

\date{\today}

\begin{abstract}
 We study the triangular lattice Ising model with a finite number of vertically stacked layers and demonstrate a low temperature reentrance of two Berezinskii-Kosterlitz-Thouless transitions, which results in an extended disordered regime down to $T=0$. Numerical results are complemented with the derivation of an effective low-temperature dimer theory. 
 Contrary to order by disorder, we present a new scenario for fluctuation-induced ordering in frustrated spin systems.  While short-range spin-spin correlations are enhanced by fluctuations, quasi-long-range ordering is precluded at low enough temperatures by proliferation of topological defects.
\end{abstract}


\maketitle

\textit{Introduction.---}%
The antiferromagnetic triangular lattice Ising model (TLIM) is the paradigmatic example of geometric frustration~\cite{Wannier1950Antiferromagnetism,Houtappel1950Order,Husimi1950Statistics}.
Despite its simplicity, the TLIM exhibits all the defining features of a highly frustrated magnet.
The extensive degeneracy of its ground state or Wannier manifold, which comprises any state without three parallel spins on the same triangle, leads to a residual entropy density $S \approx 0.323 k_B$.
This property makes the system very sensitive to perturbations, 
as is manifested in the algebraic spin-spin correlations.
Simple perturbations, such as further-neighbor couplings, relieve the frustration and induce long-range order (LRO) or quasi-LRO~\cite{Mekata1977Antiferro,Landau1983Critical,Nienhuis1984Triangular,Qian2004Triangular,Sato2013Quasi}.
The ground state degeneracy can also be lifted via the order by disorder mechanism~\cite{Villain1980order}. For instance,
a vertical 3D stacking of TLIMs produces a low-$T$ partially disordered  antiferromagnetic (PDA) phase consisting of two ordered sublattices with opposite magnetizations and the third one that remains disordered~\cite{Blankschtein1984,Coppersmith1985,Matsubara1987Frustrated,Heinonen1989Critical,Kim1990,Bunker1993Multiple-histogram}. 
By adding  a transverse field, we obtain the  quantum Ising model (QIM) that also contains a low-$T$ PDA phase stabilized by quantum fluctuations~\cite{Moessner2000Two-Dimensional,Isakov2003,Jiang2005String}.

In this Letter, 
we show an exotic classical spin liquid phase with unusual pseudocritical correlations in a simple generalization of the TLIM, namely, a vertically stacked {\it finite} number $N_z$ of triangular layers.
This new 
phase consists of a line of pseudocritical disordered states. Surprisingly, spins  become more {\it correlated} at short distances with increasing temperature: the spin correlation falls off like $\sim\hspace{-5pt} r^{-\eta(T)} e^{-r/\xi}$ with the exponentially large correlation length, $\ln \xi \propto J/T$, and the short-distance effective power law decay becomes {\it slower} at higher $T$ ($d\eta/dT <0)$~\footnote{In contrast, the single-layer system ($N_z = 1$) does not exhibit such a peculiar enhancement of short-distance correlations.}. 
This 
is similar to Villain's 
order by disorder~\cite{Villain1980order}.
However, while thermal fluctuations 
increase short-distance spin-spin correlations, hence  the stiffness of the effective field theory, quasi-LRO sets in only when the stiffness reaches a critical value necessary to suppress the proliferation of topological defects. 

Our study is in part motivated by recent advances in film-growth techniques~\cite{Vaz2008Magnetism,Leusink.arXiv1309.2441} and fabrication of artificial spin systems~\cite{Zhang11nan}. The model  Hamiltonian is
\begin{equation}
  \ham=J\sum_{n=1}^{N_z}\sum_{\langle i, j\rangle}\sigma^{z}_{i, n}\sigma^{z}_{j, n}-J_z\sum_i\sum_{n=1}^{N_z}\sigma^{z}_{i,n}\sigma^{z}_{i, n+1},~~ J>0,
  \label{eq:H}
\end{equation}
where  $\sigma^{z}_{i, n}=\pm 1$ is an Ising spin at site $i$ of the $n$th layer, and $\langle i, j\rangle$ runs over intralayer nearest-neighbor sites. 
We use open (periodic) boundary conditions along the vertical direction (in the $ab$ plane).  Although not essential, we assume a ferromagnetic (FM) interlayer exchange $J_z>0$.

We will see that the  phase diagram of $\ham$  changes with $N_z$, but there is always  an extended pseudocritical phase right above $T=0$ [Fig.~\ref{fig:phase-diagram}(b)]. The single-layer TLIM has no phase transition at any finite $T$. For $N_z>1$, while thermal fluctuations also destroy the critical state at $T = 0$, 
the configurational entropy 
enhances  
the short-range in-plane correlations, 
leading to the classical spin liquid phase.
For $N_z < N_{c1}$ the system remains disordered at any finite $T$ and the peculiar low-$T$ state 
crosses over to the high-$T$ paramagnetic (PM) state. 
For $N_{c1}\leq N_z<N_{c2}$, 
fluctuations 
induce a Berezinskii-Kosterlitz-Thouless (BKT) transition to a critical phase that is destroyed by another BKT transition at a higher temperature of order $J$.
Conversely, the lowest-$T$ BKT transition defines a {\it reentrant transition} back to the disordered low-$T$ regime.
Finally, for $N_{c2} \leq N_z < \infty$,  a PDA phase emerges in the middle of the BKT phase.

The degenerate ground states of the TLIM are related to dimer coverings~\cite{Blote1982Roughening,Nienhuis1984Triangular} and fully packed loops~\cite{Blote1994Fully} on the dual honeycomb lattice~\footnote{Because the midpoints of the kagome lattice links form a honeycomb lattice, the TLIM is also relevant for systems such as kagome spin ice~\cite{Moessner03,Moller09,Chern11} and the kagome QIM~\cite{Nikolic03}.}.
To account for the exotic low-$T$ physics, we derive a  low-energy dimer model.
Entropic effects  generate interdimer interactions and topological defects that  control the low-$T$ physics and
quantitatively reproduce the  results of our  Monte Carlo (MC) simulations of $\ham$.
The global phase diagram is also obtained by a different mapping of $\ham$ into a single-layer QIM.

\begin{figure}[t]
  \includegraphics[width=0.98\hsize,bb=0 0 1019 599]{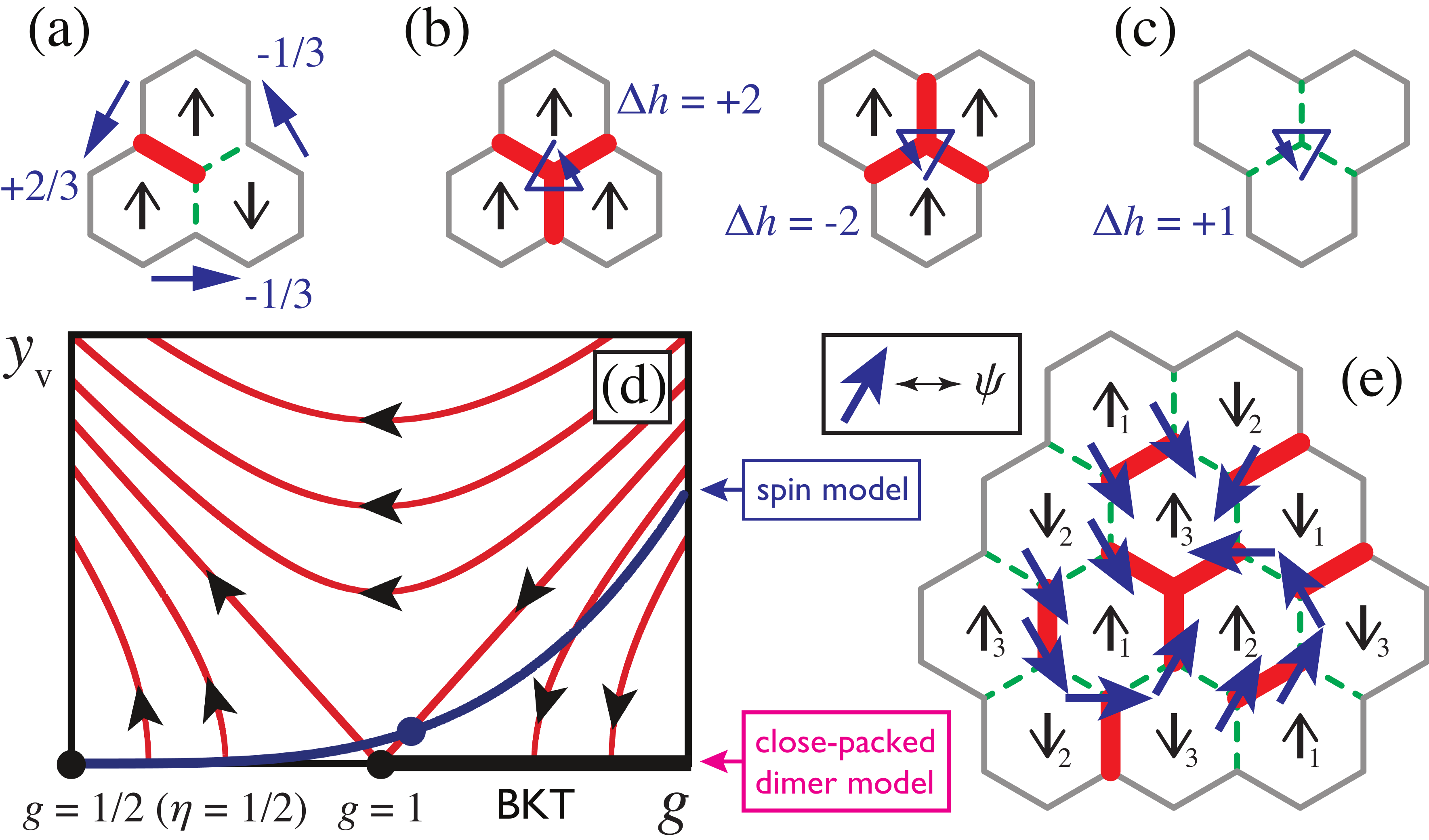}
  \caption{\label{fig:KT} 
    (a) Mapping to a dimer state (each arrow stands for the majority spin of a chain). 
    (b) Physical topological defects corresponding to height dislocations $\Delta h = \pm 2$. 
    (c) Example of a monomer, which  has no physical realization in the spin model.
    (d) Renormalization-group flow diagram near the lowest-$T$ BKT transition ($g=1$). The trajectory schematically shows the bare coupling of the spin model with $N_z > N_{c1}$, while the $y_\text{v}=0$ axis corresponds to the close-packed dimer model. 
    (e) Example of a vortex of the spin operator $\psi=\sigma^{z}_1 + \sigma^{z}_2 e^{2\pi i/3} + \sigma^{z}_3 e^{4\pi i/3}$ associated with an isolated topological defect (note that three parallel spins correspond to $\lvert\psi\rvert=0$, as expected for a vortex core).}
\end{figure}

\textit{Dimer coverings and effective low-$T$ theory.---}
The highly degenerate  ground space of $\ham$ is also a Wannier manifold because it consists of FM vertical chains, which can be treated as effective Ising variables. This manifold is conveniently mapped onto the set of dimer coverings on the dual honeycomb lattice~\cite{Blote1982Roughening,Nienhuis1984Triangular} by placing a dimer on every  link that crosses a frustrated (i.e., up-up or down-down) bond [see Fig.~\ref{fig:KT}(a)].
Because each triangle has only one frustrated bond, exactly one dimer is attached to each honeycomb lattice site.
The $T = 0$ partition function of  $\ham$  is mapped onto a  partition function  for dimers  with the same statistical weight for each dimer configuration.

By assuming $\lvert{J_z}\rvert < 2 N_z J$, we can include thermal effects into an effective single-layer dimer model 
for 
$T \ll \lvert{J_z}\rvert$.
We first consider lowest-energy excitations  that create a single kink in a chain surrounded by three-up and three-down chains.
The minimum excitation energy  is $2\lvert{J_z}\rvert$, because the interchain molecular field is zero.
Dressed by such kink excitations, these chains  ($\usebox{\dimerIIIA}$ or $\usebox{\dimerIIIB}$ in the dimer representation) acquire a 
higher statistical weight $W \approx 1 + (N_z - 1) w$, where $w = \exp(-2\lvert{J_z}\rvert/T)$ and the prefactor $N_z-1$ accounts for the possible locations of the kink along the chain.
The finite-layer TLIM is then described by an action
\begin{align}
  \mathcal{S}_\text{dimer} = -K_3 \sum_{i} \left[n_{i}\,\bigl(\usebox{\dimerIIIA}\bigr) + n_{i}\,\bigl(\usebox{\dimerIIIB}\bigr)\right] + \cdots,
  \label{eq:S}
\end{align}
where the dimer-covering constraint is implicit and $n_{i}(\cdot) = 1$ ($0$) if the plaquette $i$ 
has (does not have) a designated dimer configuration and $K_3 = (N_z - 1)w + O(w^2)$. For
simplicity, we have omitted the second order terms
[see the Supplemental Material].

Because the low-$T$ regime is close to the critical $T=0$ state, we pursue an effective field theory to study the critical properties of the above dimer model. 
Following the standard approach~\cite{Kondev1995Four-coloring,Kondev1996Kac-Moody,Alet2005Interacting,Alet2006Classical,Papanikolaou12007Quantum,Otsuka2011Monomer-Dimer,Chandra1994},
we assign a discrete height 
$h_i$ to each plaquette $i$ such that $h_i$ changes by $2/3$ ($-1/3$) when crossing a dimer (empty link) while going counterclockwise around a site of one sublattice of the honeycomb lattice 
[Fig.~\ref{fig:KT}(a)]. The dimer constraints assure a consistent height profile.

The critical spin states correspond to the roughing phase of the coarse-grained height field $h(\mathbf r)$ 
described by a Gaussian theory. Taking into account the locking potential associated with the discreteness of the  height variables, the effective long-wavelength theory is given by a standard sine-Gordon action:
\begin{align}
  \mathcal{S}_{\text{eff}} = \int d^2 \mathbf{r} \left[ \pi g \left(\nabla h\right)^2 + u^{}_p \cos\left(2p\pi h\right)\right],~~p=3.
  \label{eq:Seff}
\end{align}
Here $g$ is the stiffness and $u_{p=3}$ is 
the locking potential amplitude.
In the Coulomb gas description equivalent to $\mathcal{S}_\text{eff}$~\cite{Nienhuis1984Critical,Jose1977RG}, the locking term carries an ``electric'' charge $p$ and its scaling dimension is $\Delta_p = p^2/(2g)$. The locking potential becomes 
relevant
 for $g > 9/4$.
Due to the  periodicity in the height variable the dimer operator carries $p=1$, i.e., $\Delta_\text{dimer} = 1/(2g)$,
and we infer 
$g=1/2$ for the TLIM at $T = 0$~\footnote{This is consistent with the observation that $u_{p=3}$ is irrelevant with this stiffness.} from the exact dimer correlator~\cite{Fisher1961Statistical,Temperley1961Dimer,Kasteleyn1963Dimer}.

The fluctuation-induced dimer interaction
increases the stiffness $g$ because $K_3 > 0$ favors the columnar dimer state (flat landscape after coarse graining).
In addition, an exponentially small but finite concentration of defects 
violating
the constraint also appears at finite $T$.
The simplest example is a triangle of parallel  spin chains, which corresponds to a height dislocation $\Delta h = \pm 2$ [Fig.~\ref{fig:KT}(b)]. These defects correspond to vortices of the spin operator $\psi$~\footnote{$\psi$ can serve as an order parameter of both the PDA state and the up-up-down (or down-down-up) ferrimagnetic state.} with winding number $\pm 1$~[Fig.~\ref{fig:KT}(e)]. 
The factor of $2$ arises because 
the associated vertex operator 
has $p=1/2$, i.e., $\psi \sim \exp(i\pi h)$.
Another crucial observation is that unitary ($\Delta h = \pm 1$) dislocations, namely monomers [see Fig.~\ref{fig:KT}(c)], are not physical excitations of the spin model. Monomers are known to induce a three-state Potts transition~\cite{Otsuka2011Monomer-Dimer}. 
The absence of monomers implies that our dimer model must undergo a BKT transition before reaching the ordered state.

After introducing height dislocations 
with $\Delta h = \pm 2$, 
the effective theory becomes a two-component Coulomb gas~\cite{Nienhuis1984Critical,Jose1977RG}. The dislocations $\Delta h = \pm 2$ carry a ``magnetic'' charge $q = \pm 2$ and have scaling dimension $\Delta_\text{v} = 2g$.
Thus, 
although the bare defect fugacity $y^{}_\text{v}$ is exponentially small at low $T$,
it is a relevant perturbation that  destabilizes the critical $T=0$ correlations: $\Delta_\text{v} \approx 1 < d=2$ for $g \approx 1/2$. However, as the height field becomes stiffer with increasing $T$, 
the defect fugacity $y_v$ becomes {\it irrelevant} for $g > 1$, where
the magnetic charges form bound pairs [Fig.~\ref{fig:KT}(d)].
This is a massless BKT phase extending up to $g = 9/4$ where the locking term induces a flat (ordered) state.
Our MC simulations (discussed below) show that this is 
the case for $N_z \geq N_{c1}$.
Thus, our low-$T$ theory predicts an extended  pseudocritical regime right above $T = 0$ due to proliferation of unbounded  defect triangles.

\begin{figure}[t]
  \includegraphics[width=\hsize,bb=0 0 536 228]{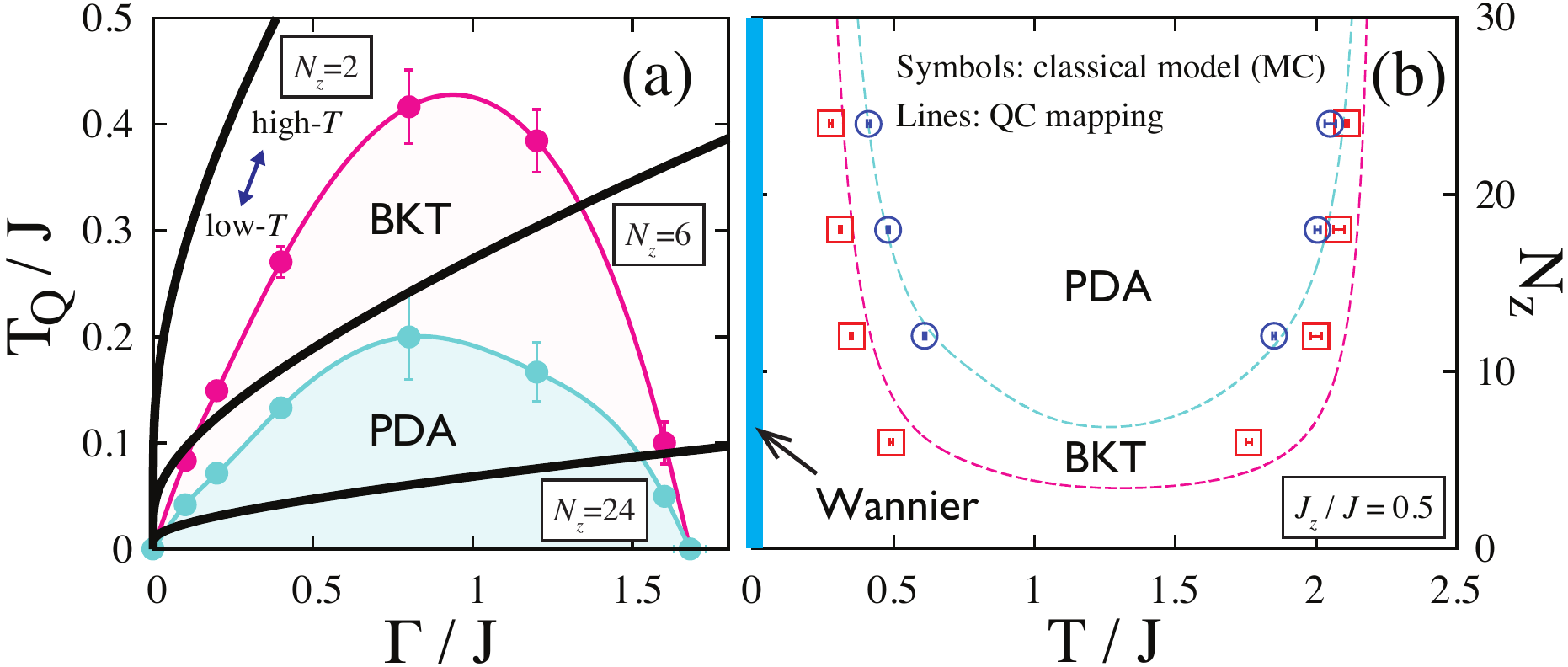}
  \caption{\label{fig:phase-diagram} 
    (a) Phase diagram of the QIM~\eqref{eq:HQ} (data: MC results from Ref.~\onlinecite{Isakov2003}) and the trajectories of quasi-2D classical systems ($J_z / J = 0.5$; $N_z = 2$, $6$, and $24$). The phase boundaries are guides to the eye. 
    (b) Phase diagram of the classical model compared with the  QC mapping.
  }
\end{figure}

\textit{Quantum-classical mapping.---}%
We can get a glimpse of the complete phase diagram of $\ham$
by using the quantum-classical (QC) correspondence. We consider the single-layer QIM:
\begin{equation}
  Z_\text{Q} = \mathrm{Tr} \exp\left(-\ham_\text{Q}/T_\text{Q}\right),~~
  \ham_\text{Q} = J\sum_{\langle{i,j}\rangle} \sigma^z_i \sigma^z_j - \Gamma \sum_i \sigma^x_i,
  \label{eq:HQ}
\end{equation}
where $\sigma^\mu_i$ are Pauli matrices at site $i$ and we use
a different symbol, $T_\text{Q}$, for the temperature of the QIM. 
The transverse field $\Gamma$  selects the PDA ground state  for $0 < \Gamma < \Gamma_c$~\cite{Isakov2003,Jiang2005String}.

The order parameter space has a sixfold clock symmetry 
corresponding to the group generated by $Z_3$ lattice rotations and $C_2$ spin rotations around the $x$ axis.
Consequently, the finite-$T_\text{Q}$ phase diagram is described by an effective 
six-state clock model~\cite{Moessner2000Two-Dimensional,Isakov2003,Jiang2005String}. 
For $0 < \Gamma < \Gamma_c$, the system undergoes two BKT transitions enclosing an intermediate critical phase with emergent U(1) symmetry~\cite{Jose1977RG} [see Fig.~\ref{fig:phase-diagram}(a)].

By discretizing the imaginary time $[0,T_\text{Q}^{-1})$ into 
$N_\tau = N_z$ slices,
the QIM is mapped to
$\ham$
 with a periodic boundary condition in the vertical direction,
whose effect becomes negligible in the large $N_z$ limit.
The mapping 
is given by $T = N_\tau T_\text{Q}$ and $J_z/T = -(1/2)\ln\tanh[\Gamma/(N_\tau T_\text{Q})]$ (see the Supplemental Material for details). While this mapping is exact only for $N_\tau \to \infty$, it is still a good approximation if $\Delta \tau \equiv T_\text{Q}^{-1}/N_\tau = T^{-1}$ is much smaller than the correlation length along the imaginary time axis $\xi_\tau$. 
In this way we obtain 
\begin{align}
  \Gamma(T, J_z) = T \tanh^{-1} \exp \left(-2J_z/T\right).
  \label{eq:mapping}
\end{align}
Thus, although $\ham_\text{Q}$ per se does not exhibit ``stiffness from disorder'' (i.e., LRO sets in at low $T_\text{Q}$), 
varying $T$ of the classical system corresponds to changing \textit{both} $T_\text{Q}$ and $\Gamma$ in the 
phase diagram of the QIM.
As is shown in Fig.~\ref{fig:phase-diagram}(a), we expect three different scenarios depending on $N_z$ and $J_z/J$ in $\ham$:
(i) four BKT transitions with massless BKT and massive PDA phases, (ii) two BKT transitions with an intermediate massless phase, and (iii) a PM state at any $T > 0$.
In particular, the disordered low-$T$ regime predicted by the dimer model is confirmed by the QC mapping.
Finally,  because $\xi_{\tau} \simeq \Gamma^{-1}$ for $T_\text{Q} \ll J$,  
Eq.~\eqref{eq:mapping} implies
that the QC mapping is only valid for $T \lesssim J_z$.

\begin{figure}
  \includegraphics[width=\hsize,bb=0 0 462 726]{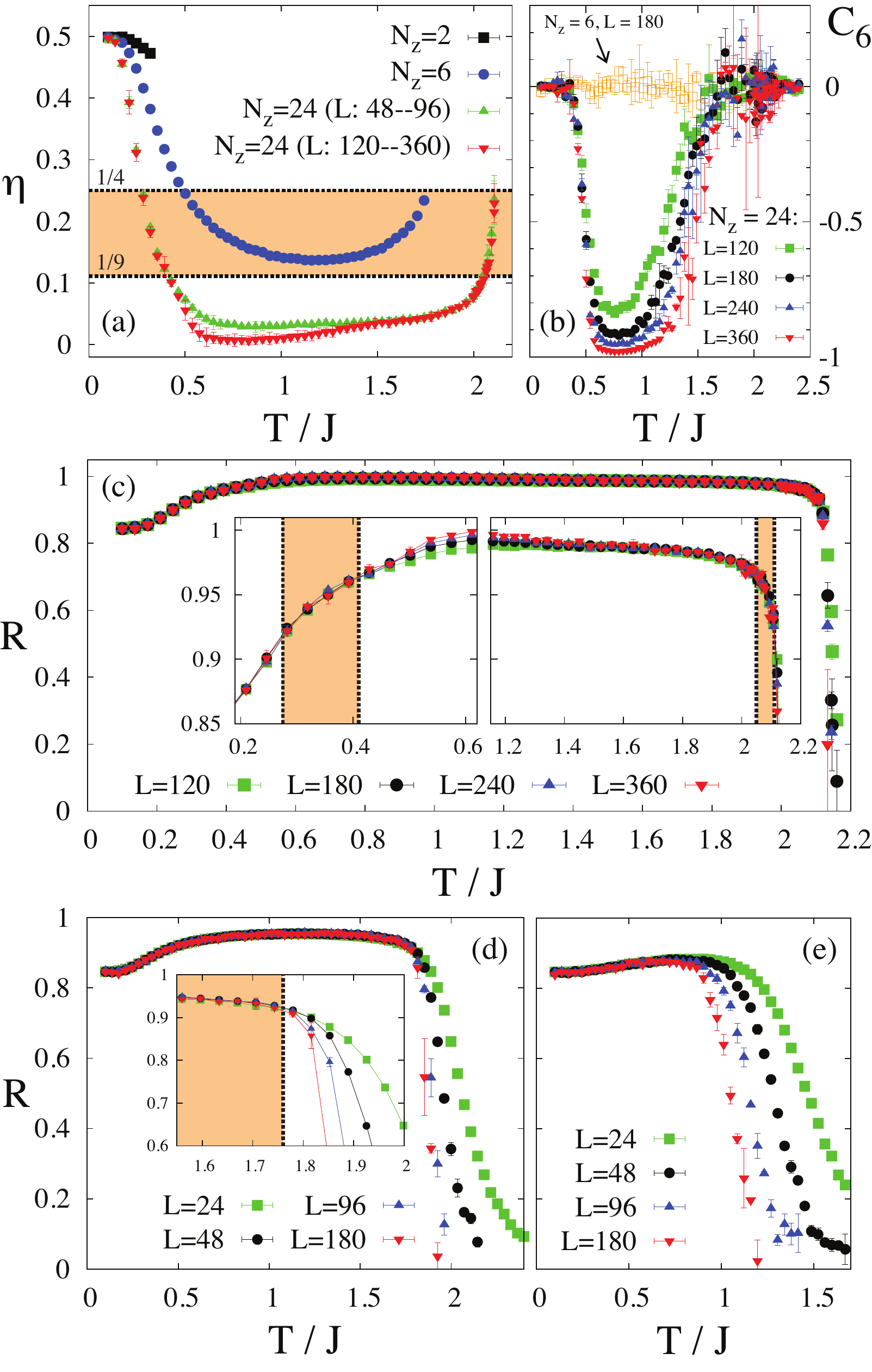}
  \caption{\label{fig:spin-MC} 
    (a) $T$ dependence of the critical exponent $\eta$.
    (b) $T$ dependence of $C_6$.
    (c)--(e) $R = G(L/2,0)/G(L/4,0)$ for $N_z = 24$, $6$, and $2$.
    The shaded regions in (a) and the insets of (c) and (d) indicate the BKT phase.
  }
\end{figure}

\textit{MC results of the spin model.---}
We confirm the above 
global phase diagram with direct MC simulations of  $\ham$ (see the Supplemental Material for details).
Below
we fix $J_z/J=0.5$ and change $N_z$ to demonstrate the scenarios (i)--(iii). 
The order parameter of the PDA state is the $\mathbf{Q} = (2\pi/3,-2\pi/3, 0)$ Fourier component of the magnetization $\Psi=\sum_{s=1}^3 M_s\,\exp[2 (s-1) \pi i/3]$, where $M_s$ is the the $s$th sublattice magnetization ($1 \le s \le 3$)~\cite{Isakov2003}. 
We also compute $C_6= {\langle \mathrm{Re}[\Psi^6]\rangle}/{\langle \lvert{\Psi}\rvert^6\rangle}$ to distinguish LRO from quasi-LRO and 
the correlation function $G(\mathbf{r})=\mathrm{Re}\,\langle\psi^*(\mathbf{r})\,\psi(0)\rangle$, where  $\psi(\mathbf{r}) = N_z^{-1} \sum_n (\sigma^{z}_{\mathbf{r}, n} +\sigma^{z}_{\mathbf{r}+{\bf e}_1, n}  e^{2 \pi i/3}+ \sigma^{z}_{\mathbf{r}+{\bf e}_2, n}  e^{4 \pi i/3} ) e^{i\mathbf{Q}\cdot\mathbf{r}}$ is the local order parameter. $C_6$ equals $-1$ ($+1$) for perfect PDA (ferrimagnetic) order~\cite{Isakov2003}.
If the system has LRO,  $R=G(L/2,0)/G(L/4,0)$ goes to unity for a lateral size $L \gg \xi$, while $R^{}\to 0$ in the PM phase. $R$ is particularly useful for detecting quasi-LRO because it becomes $L$ independent when the system is critical~\cite{Tomita2002}.

The exponent $\eta$ [see Fig.~\ref{fig:spin-MC}(a)] characterizing the spin-spin correlation function is estimated from the standard finite-size scaling hypothesis in $d = 2$ dimensions: $\lvert{\Psi}\rvert \sim L^{-\eta/2}$  (see Fig.~S3 in the Supplemental Material). 
This is a convenient quantity to locate BKT transitions because it takes a universal value. By analyzing scaling dimensions of perturbative operators that become marginal at each transition, Jose \textit{et al.}~have shown~\cite{Jose1977RG} that $\eta=1/4$ ($\eta=1/9$) at the BKT transition from the PM (PDA) state to the critical BKT phase.
For instance, for the reentrant BKT transitions, we know that  $\eta = 1/(4g)$ from the scaling dimension of $\psi$, while 
$g=1$ and $g=9/4$ for the lower and upper BKT transitions, respectively.
$\eta$ changes continuously between $1/4$ and $1/9$ in the critical phase.

Our simulation results for $N_z = 2$, $6$, and $24$ are summarized in Fig.~\ref{fig:spin-MC}, which clearly shows three distinct behaviors corresponding to the scenarios (i)--(iii). For $N_z = 24$, the ratio $R$ becomes $L$ independent in two temperature regimes in which the effective exponent $\eta$ interpolates between 1/4 and 1/9, indicating two extended critical phases. A PDA phase, corresponding to a negative $C_6$, is sandwiched by 
these
critical regimes. This LRO disappears in the $N_z = 6$ system ($\eta$ never falls below $1/9$).
Finally, 
for $N_z = 2$,
the $R(T)$ curves for different $L$ seem to merge at low temperatures. 
However, 
the corresponding temperature range decreases systematically with increasing $L$, implying a PM state at any finite $T$ [see Fig.~\ref{fig:spin-MC}(e)]. Interestingly, while $\eta_\text{eff} \gg 1/4$ confirms the PM nature at $T > 0$, the exponent approaches the $T=0$ value ($\eta = 1/2$) from {\it below}. 
This is peculiar
because the high-$T$ trivial exponent is $\eta^{}_\text{eff}=2$~\cite{Holtschneider2005Two-dimensional}, and it indicates a crossover from an unstable fixed point~\cite{Kamiya2010Crossover}. The boundaries of the ($T/J, N_z$) phase diagram shown in Fig.~\ref{fig:phase-diagram}(b)  agree quite well with the QC mapping.
A small systematic shift is caused by the different boundary conditions in the vertical direction
mentioned above.

\begin{figure}[tb]
  \includegraphics[width=\hsize,bb=0 0 346 201]{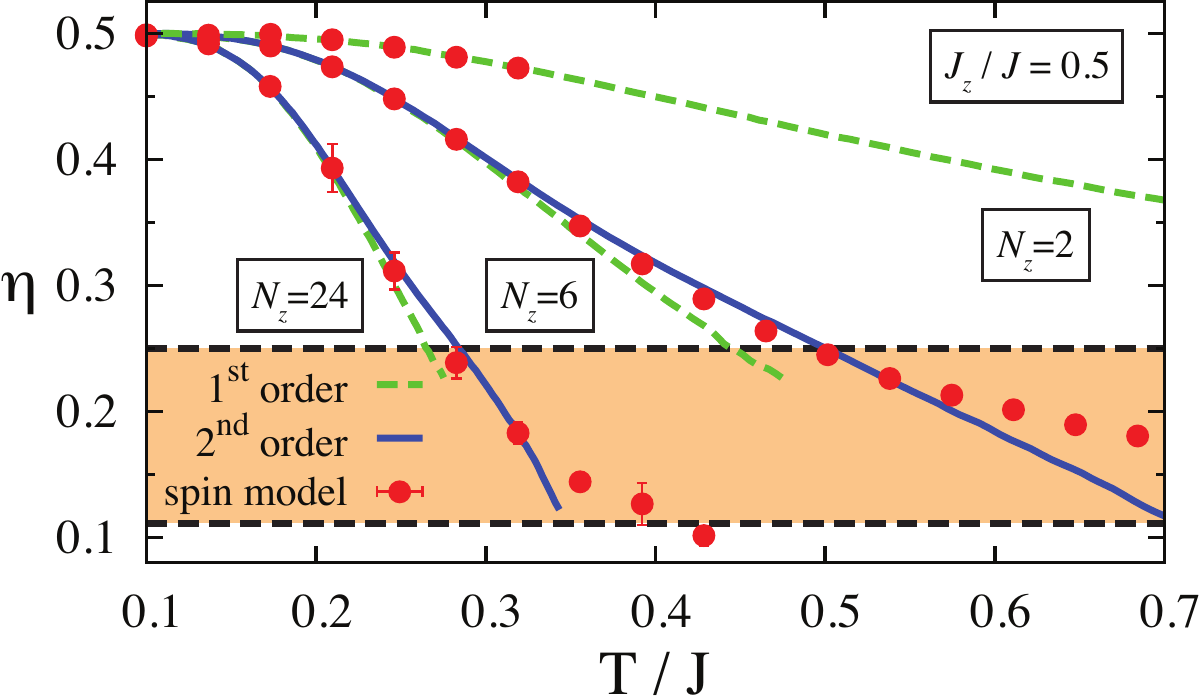}
  \caption{\label{fig:comparison} 
    Comparison of exponents in the low-$T$ regime. The shaded region corresponds to the BKT phase (the horizontal lines indicate $\eta = 1/4$ and $\eta = 1/9$).
    The dashed (solid) lines are the results of simulating the dimer model with interactions up to first (second) order in $w$ (error bars  are smaller than the line width). The points are the results of MC simulations of the spin model [Fig.~\ref{fig:spin-MC}(a)].}
\end{figure}

The puzzling low-$T$ physics can be explained with the aid of our low-energy dimer model.
By using the directed-loop MC algorithm~\cite{Sandvik2006Correlation}, we estimate the stiffness $g$ by evaluating the winding number fluctuations~\cite{Alet2005Interacting,Alet2006Classical} of the dimer model $\mathcal{S}_\text{dimer}$ (without defects) 
as a function of $T$, $N_z$, and $J_z$.
The exponent $\eta = 1/(4g)$ must coincide with the effective exponent obtained from our MC simulations in the pseudocritical regime of the spin model 
($ \langle \sigma_{i,n} \sigma_{i+r,n} \rangle \sim r^{-\eta(T)} e^{-r/\xi}$)
because $\xi \propto {y_\text{v}}^{-1}$ is exponentially large in $J/T$ and consequently much larger than $L$.
For $N_z = 6$ and $24$ we simulate both the first- [Eq.~\eqref{eq:S}] and second-order (see the Supplemental Material) effective theories, while for $N_z = 2$ we use only the first order expression because second order contributions in $w$ do not exist in this case.
The excellent agreement between these results and those obtained directly from $\ham$ (Fig.~\ref{fig:comparison})  confirms the validity of the effective low-$T$ dimer model.
The discrepancy at the lowest-$T$ BKT transition (where $\eta = 1/4$)
for $N_z = 6$ and $24$ is $\lesssim 5\%$. 
Further discrepancies above the critical temperature
indicate the breakdown of perturbation theory because  $K_3$ [Eq.~\eqref{eq:S}] becomes of order 1.

In summary, the reentrant BKT transition of the TLIM with a finite number 
of vertically stacked layers leads to a
low-$T$ pseudocritical spin liquid phase. 
Based on a renormalization-group analysis of an effective dimer model, we unveiled the ``stiffness from disorder'' phenomenon that explains 
this exotic behavior.
Our work underscores the subtle interplay between  thermal fluctuations and topological defects.
While thermal fluctuations enhance spin-spin correlations, quasi-LRO sets in only when the stiffness reaches the critical value required to suppress  proliferation of topological defects. The ubiquitous nature of the Ising model offers alternative routes for realizing this exotic low-$T$ physics. 
In particular, the multilayered TLIM described by $ \ham$ can be realized with thin films of CsCoCl$_3$~\cite{Mekata1977Antiferro,Mekata78}, buckled colloidal monolayers~\cite{Han08}, or  nanomagnet arrays~\cite{Zhang11nan}.

\noindent \textit{Acknowledgments---} 
We thank Y.~Tomita, Y.~Motome, A.~Furusaki, H.~Otsuka, H.~Mori and P.~Chandra for valuable discussions.
We also thank S.~Isakov for providing his numerical data of the QIM.
Computer resources for numerical calculations were supported by the Institutional Computing Program in LANL. This work was carried out under the auspices of the NNSA of the U.~S.~DOE at LANL under Contract No.~DE-AC52-06NA25396, and was supported by the U.~S.~DOE, Office of Basic Energy Sciences, Division of Materials Sciences and Engineering. 
Y.~K.~is grateful to the support by the RIKEN iTHES project.


\appendix

\vspace{15pt}

\begin{center}
  {\bf ---Supplemental Material---}
\end{center}

\newsavebox{\dimerpairB}
\savebox{\dimerpairB}[17.3pt][s]{%
  \setlength{\unitlength}{2.pt}
  \begin{picture}(6.928,4)(0.3,0.9)
    \linethickness{1.6pt}
    \put(1.732,4){\line(-1.732,-1){1.732}}
    \put(1.732,0){\line(-1.732,1){1.732}}
    \put(5.196,4){\line(1.732,-1){1.732}}
    \put(5.196,0){\line(1.732,1){1.732}}
    \linethickness{1.8pt}
    \put(3.464,1){\line(0,1){2}}
    \linethickness{0.05pt}
    \put(1.732,4){\line(1.732,-1){1.732}}
    \put(1.732,0){\line(1.732,1){1.732}}
    \put(0,1){\line(0,1){2}}
    \put(5.196,4){\line(-1.732,-1){1.732}}
    \put(5.196,0){\line(-1.732,1){1.732}}
    \put(6.928,1){\line(0,1){2}}
  \end{picture}
  \setlength{\unitlength}{1pt}
}

\newsavebox{\dimerpairCI}
\savebox{\dimerpairCI}[17.3pt][s]{%
  \setlength{\unitlength}{2.pt}
  \begin{picture}(6.928,4)(0.3,0.9)
    \linethickness{1.6pt}
    \put(1.732,4){\line(-1.732,-1){1.732}}
    \put(1.732,0){\line(-1.732,1){1.732}}
    \put(5.196,4){\line(-1.732,-1){1.732}}
    \put(5.196,0){\line(-1.732,1){1.732}}
    \put(6.928,1){\line(0,1){2}}
    \linethickness{0.05pt}
    \put(1.732,4){\line(1.732,-1){1.732}}
    \put(1.732,0){\line(1.732,1){1.732}}
    \put(3.464,1){\line(0,1){2}}
    \put(0,1){\line(0,1){2}}
    \put(5.196,4){\line(1.732,-1){1.732}}
    \put(5.196,0){\line(1.732,1){1.732}}
  \end{picture}
  \setlength{\unitlength}{1pt}
}

\newsavebox{\dimerpairCII}
\savebox{\dimerpairCII}[17.3pt][s]{%
  \setlength{\unitlength}{2.pt}
  \begin{picture}(6.928,4)(0.3,0.9)
    \linethickness{1.6pt}
    \put(1.732,4){\line(1.732,-1){1.732}}
    \put(1.732,0){\line(1.732,1){1.732}}
    \put(0,1){\line(0,1){2}}
    \put(5.196,4){\line(1.732,-1){1.732}}
    \put(5.196,0){\line(1.732,1){1.732}}
    \linethickness{0.05pt}
    \put(3.464,1){\line(0,1){2}}
    \put(1.732,4){\line(-1.732,-1){1.732}}
    \put(1.732,0){\line(-1.732,1){1.732}}
    \put(5.196,4){\line(-1.732,-1){1.732}}
    \put(5.196,0){\line(-1.732,1){1.732}}
    \put(6.928,1){\line(0,1){2}}
  \end{picture}
  \setlength{\unitlength}{1pt}
}

\setcounter{figure}{0}
\renewcommand{\thefigure}{S\arabic{figure}}
\setcounter{equation}{0}
\renewcommand{\theequation}{S\arabic{equation}}

\section{I.~~ THE EFFECTIVE DIMER MODEL}
Here we derive the effective dimer model introduced in the main text to describe the low-$T$ regime of TLIM with $N_z$ vertically stacked layers. 
Each chain is ferromagnetically ordered at $T = 0$. Therefore, the ground space  of $\mathpzc{H}$ can be mapped to the dimer-covering of the dual honeycomb lattice~\cite{Blote1982Roughening,Nienhuis1984Triangular}, resulting in the extensively degenerate Wannier manifold~\cite{Wannier1950Antiferromagnetism,Houtappel1950Order,Husimi1950Statistics}. 
For $\lvert{J_z}\rvert < 2N_z J$, low temperature (i.e., $T \ll \lvert{J_z}\rvert$) thermal effects can be  described with an interacting dimer model that allows for topological defects. 
The interaction terms originate from entropic fluctuations of low-energy excitations of the spin model, namely kinks that appear along chains with zero interchain molecular field. Such chains are represented by triple-dimer plaquettes, $\usebox{\dimerIIIA}$ or $\usebox{\dimerIIIB}$, in the dimer language.
The kink only costs the intrachain exchange energy of $2\lvert{J_z}\rvert$.
As long as most of their spins remain in the original state, these chains dressed by kinks can be associated to the original Wannier state.
In other words, the statistical weight factors associated with these plaquettes ($\usebox{\dimerIIIA}$ and $\usebox{\dimerIIIB}$) are enhanced in comparison to the $T = 0$ case, for which every dimer configuration has the same weight. Our ``majority rule'' leads to a small ambiguity for even $N_z$  because the chain can have no net magnetization. However, this subtlety can be properly addressed  by the perturbation theory in $N_z w$, with $w = \exp(-2\lvert{J_z}\rvert/T)$, as we explain below.

\subsection{A.~~ Single plaquette term}

\begin{figure}[tb]
  \includegraphics[width=\hsize,bb=0 0 471 309]{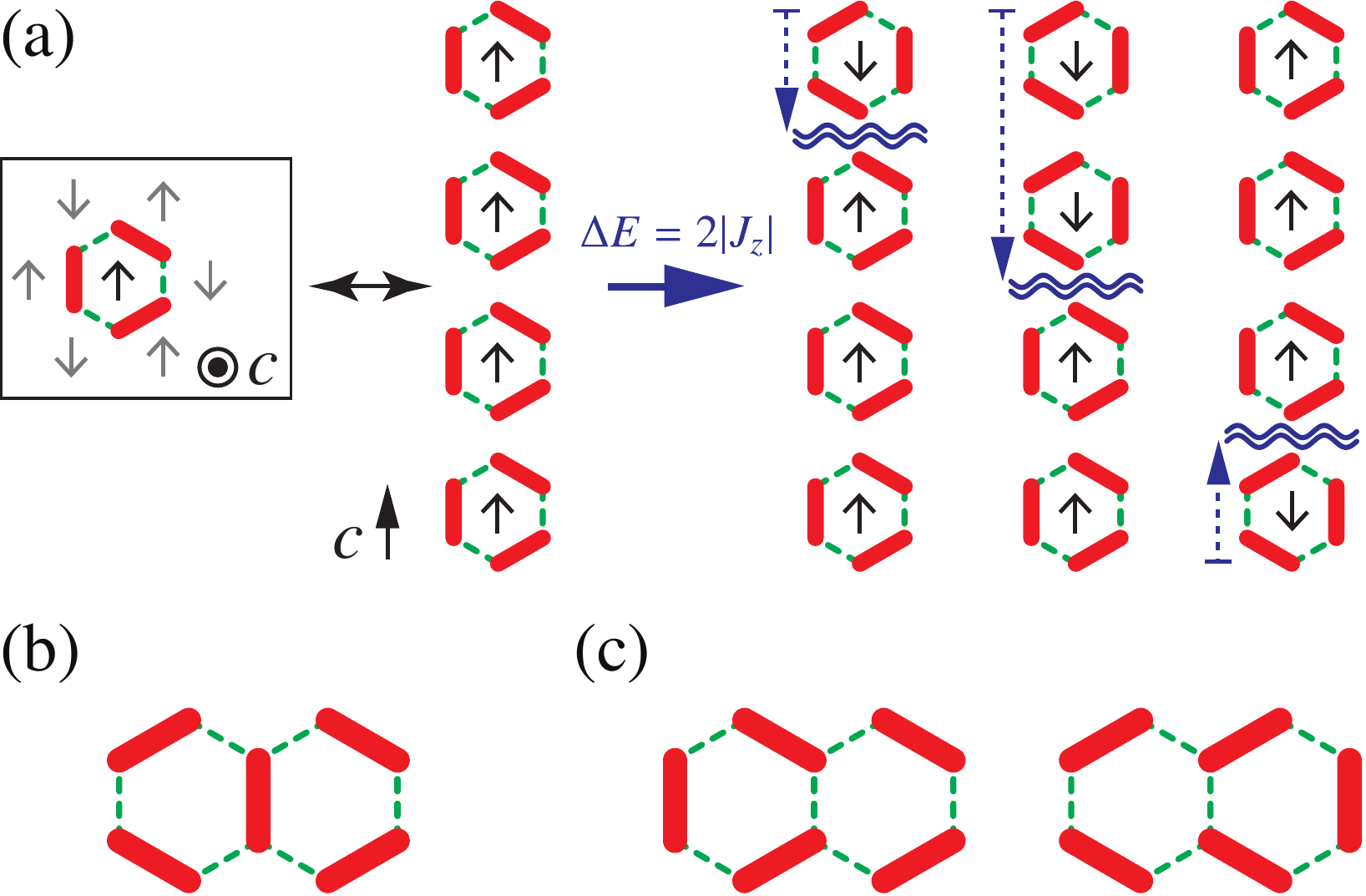}
  \caption{\label{figS:kinks}
    (a) Example of a chain ($N_z = 4$) surrounded by 3-up 3-down chains at $T=0$ (left) and the low-energy excited states obtained by inserting a kink (right).
    We depict dimers on each layer by using the same rule.
    (b) $(3,3)$ state and (c) $(3,2)$ states.
  }
\end{figure}

We first consider a single chain surrounded by 3-up 3-down chains corresponding to, say, the $\usebox{\dimerIIIA}$ plaquette.
A kink is inserted from either the top or the bottom layer to obtain first-order excited states [see Fig.~\ref{figS:kinks}(a)]~\footnote{It is interesting to note that if we define the dimer configuration on each layer and think of the vertical axis as an imaginary time dimension of a quantum model, an inserted kink corresponds to a dimer flip event.}.
We allow the kink inserted from the top layer to move up to the midpoint of the chain, while the one inserted from the bottom layer can only move up to the site that is right before the midpoint of the chain. In this way, we avoid double-counting states that are attributed to the $\usebox{\dimerIIIB}$-type plaquette.
Thus, the total number of possible kink positions associated with the original state is $N_z - 1$.

The second-order excited states contributing to the single-plaquette term are obtained by creating a pair of kinks along a single chain.
By starting from a spin chain configuration corresponding to $\usebox{\dimerIIIA}$, $(N_z - 1)(N_z - 2) / 2$ excited states are generated and the same is true by starting from the $\usebox{\dimerIIIB}$-type chain. 
The bottom line is that $(N_z - 1)(N_z - 2) / 2$ configurations should be attributed as excited states to each of $\usebox{\dimerIIIA}$ and $\usebox{\dimerIIIB}$ plaquettes.

These entropic contributions can be described by the $K_3$ term defined in Eq.~(2) of the main text.
By solving 
$\exp(K_3) = W_3 \equiv 1 + (N_z - 1)w + \frac{1}{2}(N_z - 1)(N_z - 2) w^2 + O(w^3)$,
we obtain
\begin{align}
  K_3 = (N_z - 1) w - \frac{1}{2}(N_z - 1)w^2 + O(w^3).
\end{align}

\subsection{B.~~ Two-plaquette terms}
There are nontrivial $O(w^2)$ contributions when chains of the type $\usebox{\dimerIIIA}$ or $\usebox{\dimerIIIB}$ are next to each other by sharing a dimer, as is shown in Fig.~\ref{figS:kinks}(b).
We refer to such a configuration as the ``$(3,3)$'' state.
The reason for the $O(w^2)$ correction is that a kink inserted along one chain changes the interchain molecular field of the other, and therefore the kinks cannot be placed independently.
A similar correction arises also for different chain-pair states obtained by flipping all spins along one of the two chains in the $(3,3)$ state [see Fig.~\ref{figS:kinks}(c)].
We refer to these configurations as ``$(3,2)$'' states.

We first consider the $(3,2)$ states.
Apart from the  already discussed first- and second-order excited states, these states admit the following excitations creating one kink per chain.
The first kink is inserted into the chain that corresponds to the triple-dimer plaquette (where the interchain molecular field is zero) from the top layer, and
it  can move up to the midpoint of the chain.
Because of the first kink insertion, spins on neighboring chain segments become flippable (with the minimum energy cost) from the top layer up to the position of the first kink.
In other words, the second kink may move from the top layer by flipping up to $M$ spins where $M$ is the number of spins flipped by the first kink.
Similar states can be created by inserting kinks from the bottom layer, but we need to exclude one configuration to avoid double-counting  when we interchange the plaquettes.
Hence, the total number of such kink-pair positions is $2 \sum_{M=1}^{N_z/2}M - 1 = N_z(N_z+2)/4 -1$. 
The corresponding statistical weight associated with the $(3,2)$ state is $W_{3,2} = W_3 + [N_z(N_z+2)/4 -1]w^2 + O(w^3)$.

For the $(3,3)$ states, nontrivial second-order excitations are obtained by inserting a kink into each chain from opposite ends.
To avoid double-counting of states that are already assigned to the $(3,2)$ states, we allow both kinks to move up to the site that is right before the midpoint of the chain.
By including contributions due to single-chain excitations discussed previously, $2(W_3 - 1)$, the statistical weight associated with the $(3,3)$ state is $W_{3,3} = 1 + 2(W_3 - 1) + 2(N_z/2- 1)^2 w^2 + O(w^3)$.

To account for $W_{3,2}$ and $W_{3,3}$, we need to include additional two-plaquette terms in the effective action:
\begin{align}
  \mathcal{S}_2= 
  \sum_{\langle{i,j}\rangle} 
  \left[
    K_{3,2}
    \left(
    n_{ij}\,\bigl(\usebox{\dimerpairCI}\bigr)
    + n_{ij}\,\bigl(\usebox{\dimerpairCII}\bigr)
    \right) 
    - K_{3,3}\, n_{ij}\,\bigl(\usebox{\dimerpairB}\bigr)
    \right],
\end{align}
where $\langle i, j\rangle$ runs over pairs of nearest-neighbor plaquettes
and $n_{ij}(\cdot\,\cdot) = 1$ ($0$) if the plaquette pair $i,j$ has (does not have) a designated two-plaquette dimer configuration, 
where the left (right) plaquette corresponds to plaquette $i$ ($j$).
By solving $\exp(K_{3} + K_{3,2}) = W_{3,2}$ and $\exp(2K_{3} - K_{3,3}) = W_{3,3}$, we obtain
\begin{align}
  K_{3,2} = \left[\frac{N_z(N_z + 2)}{4} - 1\right]w^2 + O(w^3),
  \\
  K_{3,3} = \frac{1}{2}(N_z^2 - 1)w^2 + O(w^3).
\end{align}

\section{II.~~ QUANTUM-CLASSICAL MAPPING}
Below we summarize the standard quantum-classical mapping that relates the quantum Ising model on the single triangular lattice layer at finite temperature $T_\text{Q}$ [see Eq.~(4)] with the classical Ising model on the stacked triangular lattice layers.
We consider the path integral representation of the partition function $Z_\text{Q}$ by discretizing the imaginary time interval $[0, T_\text{Q}^{-1})$ into $N_\tau$ time slices with an equal interval
\begin{align}
  \epsilon = \frac{T_\text{Q}^{-1}}{N_\tau},
  \label{eq:epsilon}
\end{align}
which will be regarded as the inverse temperature $T^{-1}$ of the classical system. We use
\begin{align}
  \exp(-\mathpzc{H}_\text{Q}/T_\text{Q}) 
  = \left[\exp(-\epsilon\mathpzc{H}_\text{diag}) \exp(-\epsilon\mathpzc{H}_\text{off-d})\right]^{N_\tau} + O(\epsilon),
\end{align}
where $\mathpzc{H}_\text{diag} = J \sum_{\langle{i,j}\rangle} \sigma^z_i \sigma^z_j$ and $\mathpzc{H}_\text{off-d} = - \Gamma \sum_i \sigma^x_i$ are the diagonal and off-diagonal parts of the Hamiltonian, respectively.
We write the spin configuration of the $n$-th time slice as
$\lvert{\chi^{}_n}\rangle = \prod_{\otimes\, i} \lvert{ \sigma^z_{i,n} }\rangle$,
and we insert a complete basis set at each interval:
\begin{align}
  Z_\text{Q} 
  &\approx
  \sum_{\{\chi^{}_{1 \le n \le N_\tau}\}}
  \prod_{n=1}^{N_\tau}
  \langle{\chi^{}_{n}}\rvert \exp(-\epsilon\mathpzc{H}_\text{diag}) \exp(-\epsilon\mathpzc{H}_\text{off-d}) \rvert{\chi^{}_{n-1}}\rangle
  \notag\\
  &=
  \sum_{\{\chi^{}_{1 \le n \le N_\tau}\}}
  \prod_{n=1}^{N_\tau}
  \langle{\chi^{}_{n}}\rvert \exp(-\epsilon\mathpzc{H}_\text{diag}) \rvert{\chi^{}_{n}}\rangle
  \langle{\chi^{}_{n}}\rvert
  \exp(-\epsilon\mathpzc{H}_\text{off-d}) \rvert{\chi^{}_{n-1}}\rangle
\end{align}
with the periodic boundary condition along the imaginary time axis $\rvert{\chi^{}_{0}}\rangle \equiv \rvert{\chi^{}_{N_\tau}}\rangle$.
We find
\begin{align}
  \langle{\chi^{}_{n}}\rvert \exp(-\epsilon\mathpzc{H}_\text{diag}) \rvert{\chi^{}_{n}}\rangle
  = \exp \left(-\epsilon J\sum_{\langle{ij}\rangle} \sigma^z_{i,n} \sigma^z_{j,n}\right)
\end{align}
and
\begin{align}
  &\langle{\chi^{}_{n}}\rvert \exp(-\epsilon\mathpzc{H}_\text{off-d}) \rvert{\chi^{}_{n-1}}\rangle
  =
  \prod_i 
  \langle{\sigma^{z}_{i,n}}\rvert
  \exp(\epsilon\Gamma\sigma_i^x)
  \rvert{\sigma^{z}_{i,n-1}}\rangle
  \notag\\
  &=  \prod_i 
  \sqrt{
    \sinh \epsilon\Gamma
    \cosh \epsilon\Gamma
  }
  \exp \left[
    \left(
    -\frac{1}{2}
    \ln \tanh \epsilon\Gamma
    \right)
    \sigma^{z}_{i,n} \sigma^{z}_{i,n-1}
    \right].
\end{align}
Hence, by defining
\begin{align}
  \epsilon J_z = -\frac{1}{2} \ln \tanh \epsilon\Gamma
  \label{eq:Jz}
\end{align}
and dropping unimportant proportionality constants, we obtain
\begin{align}
  Z_\text{Q} \approx \sum_{\{\sigma^z_{i,n}\}} 
  \exp\left(
  -\epsilon J   \sum_{\langle{ij}\rangle,\,n} \sigma^z_{i,n} \sigma^z_{j,n}
  +\epsilon J_z \sum_{i, n} \sigma^{z}_{i,n} \sigma^{z}_{i,n-1}
  \right).
\end{align}
This is a partition function of the classical Ising model on the stacked triangular layers. The first term represents the intralayer coupling and the second term represents the interlayer coupling.
\eqref{eq:epsilon} and \eqref{eq:Jz} are the main results referred to in the main text.

\begin{figure}
  \includegraphics[width=0.85\hsize,bb=0 0 752 572]{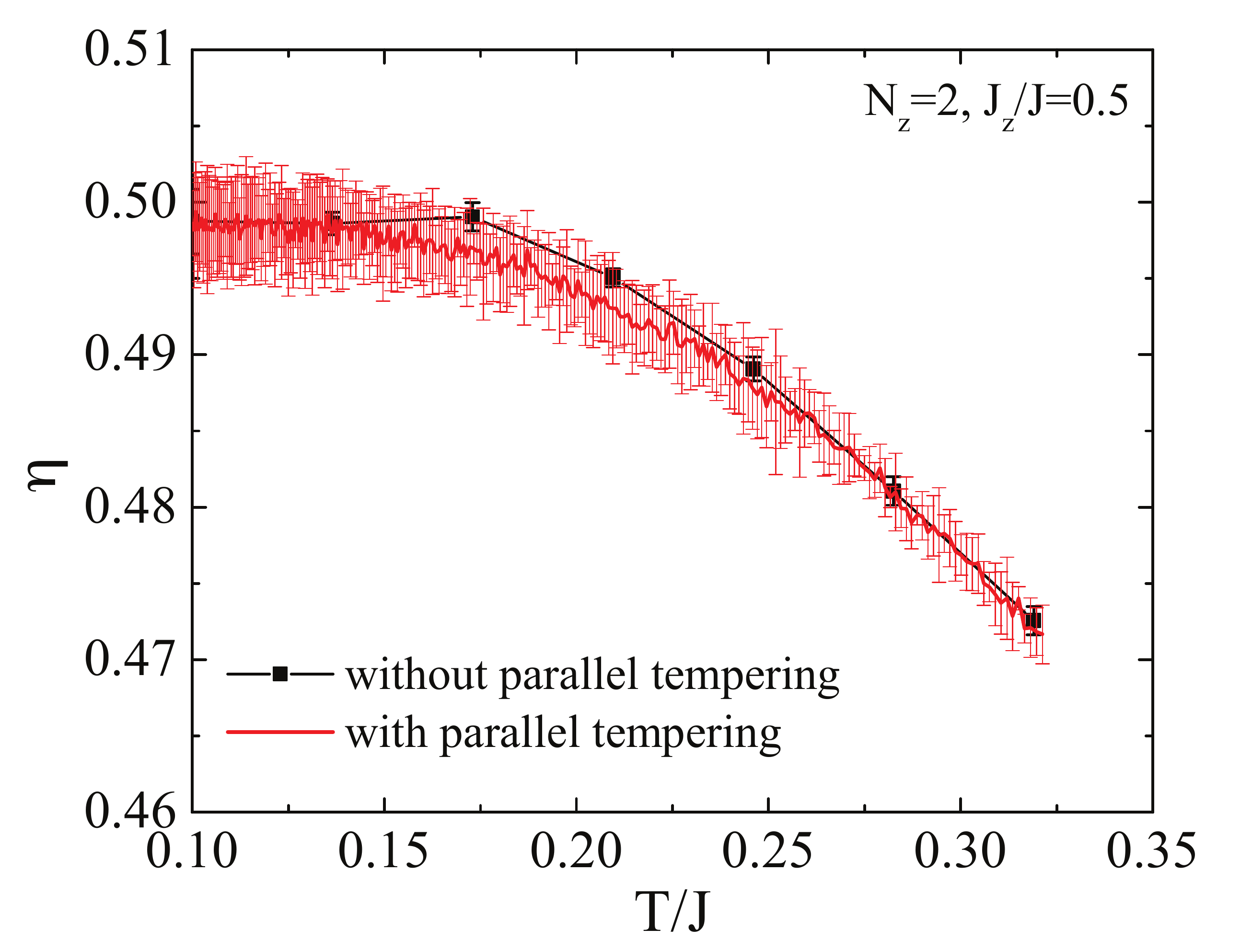}
  \caption{\label{fs2}
    Comparison of the exponent $\eta$ obtained from 
    the simulations with and without the parallel tempering.}
\end{figure}

\section{III.~~ NUMERICAL DETAILS}
In most simulations, we used $10^5$ Monte Carlo sweeps (MCS) based on the chain-cluster update method for thermalization and another $10^5$ MCS for measurements, where the measurements were done every $5$ MCS. 
In each update, we create a cluster along a chain based on the Wolff algorithm~\cite{Wolff1989collective} and then make a cluster-flip attempt by taking into account the interchain molecular field.
The system reaches equilibrium after $10^5$ MCS for the largest system size that we simulated ($L=360$ and $N_z=24$) near the BKT phase transition, $T / J = 1.9$.  To estimate the error bar, we used $6$ independent runs in most cases. 
The chain-cluster updates are useful for avoiding a freezing problem in the single-spin flip method, where the spin chains running along the $c$-axis become rather ``rigid'' for $T \ll J_z$ (the flip probability is exponentially small in $J_z/T$).
In our method, after a cluster extending along the $c$-axis is determined, the cluster flip probability is determined by the \textit{interchain} molecular field, which can be zero.

The other source of numerical difficulty is related to the fact that the system is massively degenerate at $T=0$. 
The local cluster algorithm introduced above may not explore the whole phase space. 
As a sanity check, we performed additional simulations for $N_z=2$ by using the parallel tempering method~\cite{Hukushima1996Exchange} in addition to the chain-cluster method.
As shown in Fig.~\ref{fs2}, both methods yield consistent results within the error bars, indicating that our local algorithm is efficient enough to perform a valid sampling.

\begin{figure*}
  \includegraphics[width=\hsize,bb=7 8 1244 335]{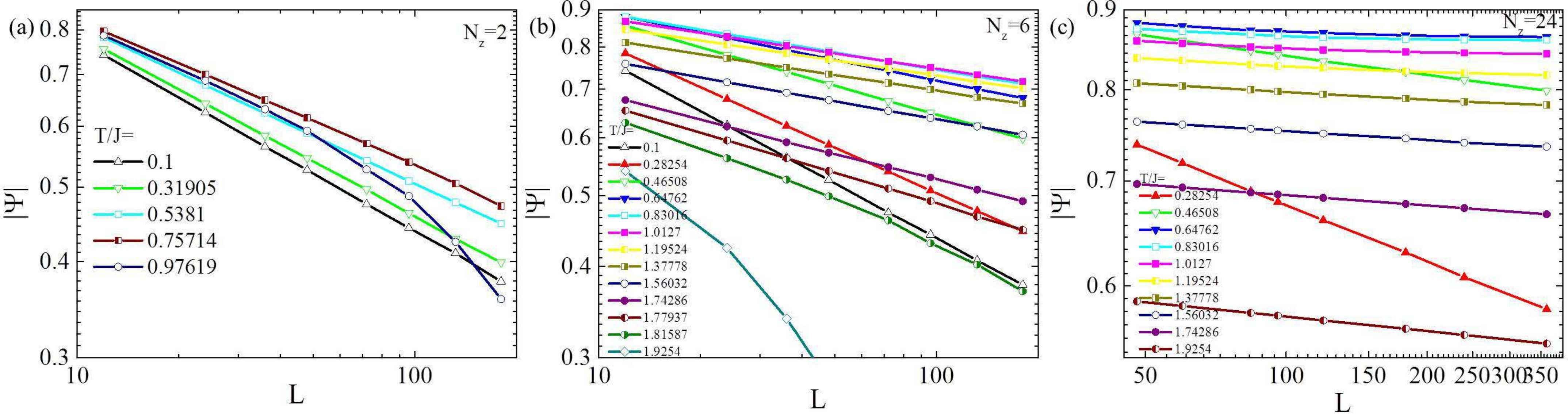}
  \caption{\label{fs3}
    $L$-dependence of $\lvert{\Psi}\rvert$ for $J_z/J=0.5$. The error bar is smaller than the symbol size. The lines are guide to eyes.
  }
\end{figure*}

\section{III.~~ DIFFERENT SCENARIOS OF PHASE TRANSITION}
Here we discuss in detail the three distinct sequences of thermodynamic transformations depending on the number of layers $N_z$. $J_z / J = 0.5$ is fixed as in the main text.

\textit{(i) Four BKT transitions.---}%
We start with $N_z=24$. The largest system size is $L=360$.
Figure~3(c) (in the main text) shows that $R(T)$ becomes $L$ independent over a finite interval below $T/J \approx 2.1$,
indicating the presence of an extended critical regime~\footnote{
  We used \unexpanded{$R=G(L/2,0)/G(L/4,0)$} to locate the BKT phases for \unexpanded{$N_z=2$} and \unexpanded{$6$} and \unexpanded{$\widetilde{R}=\widetilde{G}(L/2,0)/\widetilde{G}(L/4,0)$}, with \unexpanded{$\widetilde{G}(\mathbf{r}) = \lvert{\langle\psi^*(\mathbf{r})\,\psi(0)}\rangle\rvert$}, for \unexpanded{$N_z = 24$}, without any discernible error in the latter case.
}.
As shown in Fig.~3(a), $\eta$ decreases continuously from the expected $\eta \sim 1/4$ value for the first BKT transition at $T_\text{H}^{\eta=1/4}/J=2.13(2)$.
Here, we used finite-size scaling of the amplitude of the order parameter $|\Psi| \sim L^{-\eta/2}$. The dependence of $|\Psi|$ on $L$ is shown in Figs.~\ref{fs3}(a)--\ref{fs3}(c). 
It is well-known that a system with sixfold clock symmetry undergoes two BKT transitions in $d=2$~\cite{Jose1977RG}. 
The system then remains critical over a finite window below $T_\text{H}^{\eta=1/4}$  until the clock LRO sets in at a lower BKT transition temperature $T_\text{H}^{\eta=1/9}/J=2.05(1)$. 
Although size effects are severe for $T \sim T_\text{H}^{\eta=1/9}$, the behavior of $\eta(T)$ below $T_\text{H}^{\eta=1/9}$ is consistent with the onset of a massive phase. We find $\eta^{} \to 0$ as $L$ increases as can be seen in Fig.~3(a), and, equivalently, $|\Psi|$ saturates to a constant for large enough values of $L$, as is shown in Fig.~\ref{fs3}(c) for $N_z=24$. The $L$-dependence of $R(T)$, at least for $0.5 \lesssim T/J \lesssim 1.5$, provides further confirmation [see the insets of Fig.~3(c)].
In addition, the negative value of $C_6$ implies that the massive phase is a PDA state [see Fig.~3(b)]. 
The  $R(T)$ curves for different $L$ values  collapse again  into a single curve at lower temperatures, and the exponent falls back into the interval $1/9 \le \eta \le 1/4$.  $C_6$ vanishes below the third BKT transition temperature, $T_\text{L}^{\eta=1/9}/J=0.41(1)$, at which the exponent $\eta=1/9$ is restored [see Fig.~3(b)]. 
Below $T = T_\text{L}^{\eta=1/9}$, $\eta$ varies continuously, as is expected for a line of critical states.
We estimate the fourth BKT transition temperature as $T_\text{L}^{\eta=1/4}/J=0.275(6)$. 
The exponent seems to keep increasing continuously below $T=T_\text{L}^{\eta=1/4}$  up to the expected $T=0$ value of $\eta = 1/2$  ~\cite{Wannier1950Antiferromagnetism,Houtappel1950Order,Husimi1950Statistics} [Fig.~3(a)]. 
This result implies that $\xi$ is finite but larger than $L$. Therefore, we observe an effective exponent describing the short-range algebraic part of the spin correlations in a pseudocritical disordered phase.

\textit{(ii) Two BKT transitions.---}%
We next discuss the $N_z = 6$ case.
The main difference relative to $N_z = 24$ is the absence of LRO.
Figure~3(d) shows that the $R(T)$ curves for different $L$ merge below $T/J \approx 1.76$, suggesting a BKT transition from the high-$T$ PM phase. By analyzing $\eta$ from $\lvert{\Psi}\rvert$, we estimate $T_{\text{H}}^{\eta=1/4}/J = 1.76(2)$. 
According to Fig.~3(a), $\eta(T)$ has  a non-monotonic behavior as $T$ is further decreased.
Its minimum value ($\eta \approx 0.137$) falls into the interval $1/9 \le \eta \le 1/4$ and $C_6$ vanishes in the whole $T$-regime [Fig.~3(b)], indicating the existence of
 a massless  phase between $T_{\text{H}}^{\eta=1/4}$ and  a second BKT transition at $T_\text{L}^{\eta=1/4}/J =0.491(6)$, below which $\eta$  exceeds $1/4$ again.  
The behavior for $T \lesssim T_\text{L}^{\eta=1/4}$ is similar to the $N_z = 24$ case.

\textit{(iii) PM state at any $T > 0$.---}%
For the double layer system ($N_z = 2$),  the $R(T)$ curves for different $L$ seem to merge at low temperatures, but the corresponding temperature range decreases systematically  with increasing $L$, implying a PM state at any finite $T$~[see Fig.~3(e)]. 
However, the non-monotonic $T$ dependence of $R(T)$ as well as the effective exponent $\eta^{}$ shown in Fig.~3(a) approaching $\eta = 1/2$ at $T=0$ from below, which are not seen for $N_z = 1$ (not shown), point to the pseudocritical behavior due to the kink excitation.

\begin{figure*}
  \includegraphics[width=0.75\hsize,bb=7 5 845 335]{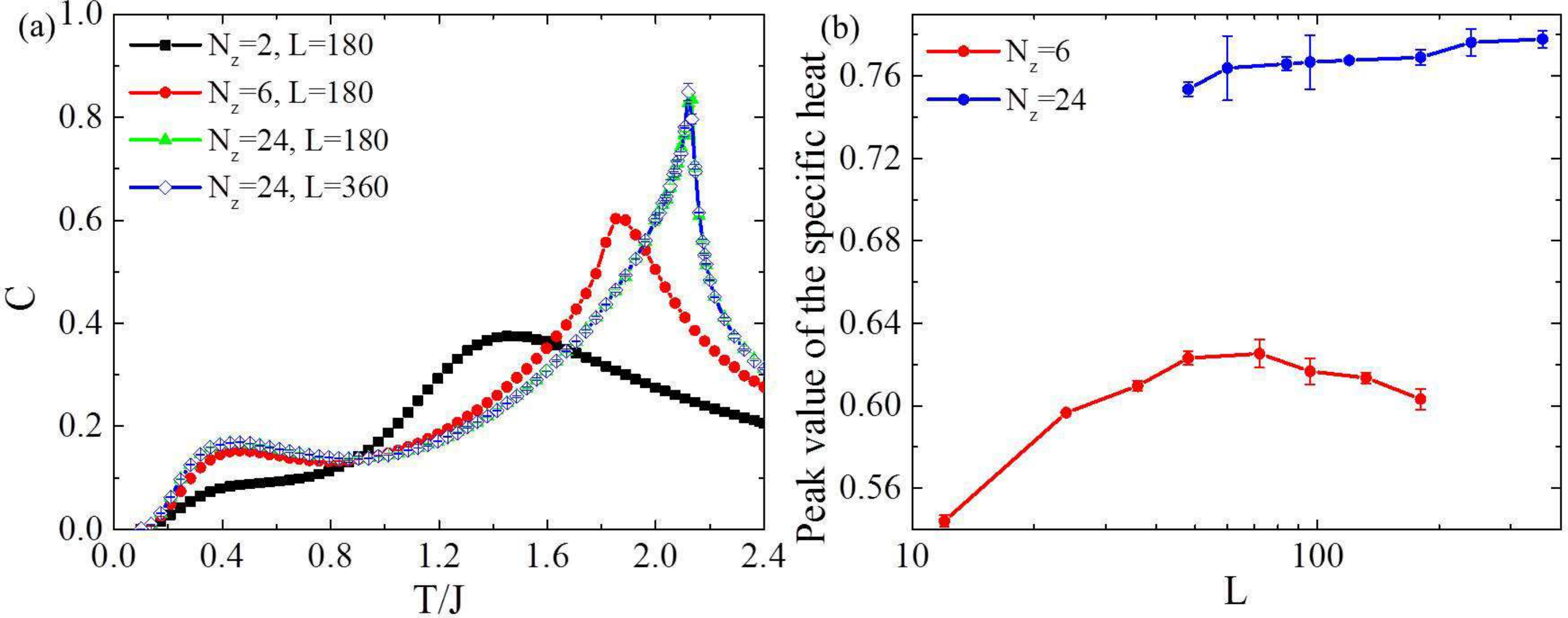}
  \caption{\label{fs4}
    (a) Specific heat $C(T)$ for $N_z=2$, $6$, and $24$. 
    (b) $L$-dependence of the peak value of the specific heat for $N_z=6$ and $N_z=24$.  
  }
\end{figure*}

\section{Specific Heat}
Figure.~\ref{fs4} shows the specific heat curves, $C=[\langle\mathpzc{H}^2\rangle-\langle\mathpzc{H}\rangle^2]/(T^2 L^2N_z)$, for different values of $N_z$. 
A hump appears around $T\approx J_z$ for the systems that undergo  BKT transitions ($N_z=6$ and $N_z=24$). 
As discussed in Refs.~\onlinecite{Coppersmith1985,Kim1990}, this hump is produced by chain excitations. 
The specific heat becomes  exponentially small for $T\ll J_z$.
The peak of $C(T)$ for $N_z = 6$ and $N_z=24$ is produced by the high temperature BKT transition. 
As expected for a BKT transition, the specific heat is almost size independent [see Fig.~\ref{fs4}(b)].

\bibliography{reference}
\end{document}